\documentclass[default,iicol]{sn-jnl}% Default with double column layout
\usepackage{graphicx}%
\usepackage{multirow}%
\usepackage{amsmath,amssymb,amsfonts}%
\usepackage{amsthm}%
\usepackage{mathrsfs}%
\usepackage[title]{appendix}%
\usepackage{xcolor, soul}%
\usepackage{textcomp}%
\usepackage{manyfoot}%
\usepackage{booktabs}%
\usepackage{mathtools}
\usepackage{lipsum}
\usepackage{algpseudocode}%
\usepackage{listings}%
\usepackage {appendix}
\begin{document}

\title[Cosmological Distances And Hubble Tension In Einstein-Cartan Theory]{Cosmological Distances And Hubble Tension In Einstein-Cartan Theory}

\author*[1]{\fnm{Siamak} \sur{Akhshabi}}\email{s.akhshabi@gu.ac.ir}

\author[2]{\fnm{Saboura} \sur{Zamani}}\email{saboura.zamani@phd.usz.edu.pl}

\affil*[1]{\orgdiv{Department of Physics, Faculty of Sciences}, \orgname{Golestan University}, \city{Gorgan}, \country{Iran}}

\affil[2]{\orgdiv{Institute of Physics}, \orgname{University of Szczecin}, \orgaddress{\street{Wielkopolska 15, 70-451}, \city{Szczecin}, \country{Poland}}}

\abstract{We analyze the measurement of cosmological distances in the presence of torsion in both Einstein-Cartan and Poincar{\'e} gauge theory of gravity. Using the modified cosmological distance measurements, we use the observed time delays in gravitational lensing systems to determine the Hubble parameter. The results show the measured Hubble parameter from a lensing system can be less than its expected value in General Relativity for certain models of torsion and its associated density parameter. This can reduce the tension between late-time and early-universe measurements of the Hubble parameter, the so-called Hubble tension.}


\maketitle
\section{Introduction}\label{Sec:Intro}
The present day Hubble parameter $H_0$ can be measured by the analysis of early universe data, \textit{i.e.}, cosmic microwave background (CMB) \cite{Planck} or by studying local or late universe phenomena, mainly gravitational lensing systems \cite{Wong, Shajib}, Cepheid variable observations \cite{Reiss1, Pietrz, Reid,  Reiss2} or using other types of stars like the red giants \cite{Freeman, Yuan, Huang}. The apparent discrepancy between these two methods, which can not be explained by measurement errors or random chance, is widely known as the Hubble tension problem \cite{Dainotti1, Dainotti2, Vagnozzi}. As an illustrating example of the difference between measured values of $H_0$ by the two methods, note that the Planck2018 data for the cosmic microwave background gives the value of $H_0$ as  $67.4 \pm 0.5$ km\,s${}^{-1}$Mpc${}^{-1}$ \cite{Planck} while according to the latest results from the SH0ES team derived from Cepheids observations of 42 nearby galaxies $H_{0}=73.0 \pm 1.4$ km\,s${}^{-1}$Mpc${}^{-1}$ \cite{Reiss2}.

The Hubble tension problem may indicates physics beyond standard models of cosmology or particle physics. Either there exists exotic types of matter like early dark energy \cite{Karwal, Evslin, Poulin, Kamin}, interacting phantom-like dark energy \cite{Valentino1, Valentino2, Joudaki} or some new relic particles \cite{Racine, Lancaster, Eramo, Kreisch} in the Universe that may solve the tension or the theory of General Relativity (GR) which is the basis for the $\Lambda$CDM model should be modified. The latter can be done in various different ways. Some authors focus on the modification to gravity at the early universe, mainly time of recombination \cite{Lin}. Others propose to modify the gravity at the weak-field limit, these include the so-called $\Lambda$-gravity model in which the cosmological constant $\Lambda$ enters into the Poisson equation of Newtonian gravity \cite{Gurzadyan1, Gurzadyan2}. Other possible approaches include scalar-tensor \cite{Ballardini, Braglia, Sola, Adi}, Horndeski theories and $f(R)$ gravity \cite{Zuma, Odintsov, Wang, Schiavone} and $f(T)$ theories \cite{Nunes}. For a review of the various methods for alleviating the Hubble tension see \cite{Knox, Valentino3} and references therein.

Among the most natural modifications to the theory of General Relativity are theories containing torsion \cite{Schrek, Hehl1, Hehl2, Shapiro}. Motivated in part by the gauge principle present in theories describing other fundamental interactions, torsion theories of gravity incorporate spin of the matter as a source for gravitational interactions \cite{Blag, Hayashi}. The geometry of spacetime in these theories is no longer that of General Relativity, \textit{i.e.} a Riemannian spacetime with a symmetric connection, but the more general Riemann-Cartan spacetime where the connection is no longer symmetric. The anti-symmetric part of the connection, called the torsion tensor, and its source, the spin density tensor of the matter fields then naturally enter the field equations and alter the gravitational dynamics and also cosmological equations. In the most general case of these theories, Poincar{\'e} gauge theory of gravity (PGT), the Lagrangian is quadratic in torsion and curvature, and there are two field equations, obtained by varying the Lagrangian with respect to dynamical variables, spin connection and tetrad fields \cite{Blag2}. The special cases of the Poincar{\'e} gauge theory include General Relativity where the torsion vanishes, Teleparallel gravity with vanishing curvature and Einstein-Cartan  (EC) gravity \cite{Kerlick, Rauch, Trautman} where the Lagrangian is the same as the Einstein-Hilbert Lagrangian of General Relativity.

From a cosmological viewpoint, the presence of torsion alters the cosmological dynamics in both early and late universe \cite{Kranas, Pereira, Medina}. Nonetheless, the effects of torsion and spin are thought to be the most pronounced in the very early universe when the spin density of the matter was very high and quantum effects play an important role. In any case, the altered expansion history in the presence of torsion causes a change to the value of present day cosmological parameters, most importantly cosmological distances, distance-redshift and Hubble-redshift relations, and the present day Hubble parameter. This may provide a way to ease the Hubble tension problem.

Recently in a series of papers, the H0LICOW Collaboration used the date from time-delay cosmography to constrain the Hubble parameter \cite{Suyu1, Sluse1, Rusu, Birrer, Sluse2, Wong}. In their latest work, they used six different lensing systems to determine $H_0$ for various cosmological models \cite{Wong}. Their results show a $5.3 \sigma$ tension between combined time-delay cosmography and the distance ladder results and Planck2018 data for the cosmic microwave background. The cosmological models analyzed in H0LICOW Collaboration works are $\Lambda$CDM, $w$CDM where the dark energy is time dependent with a equation-of-state parameter $w$ and $w_{0}w_{a}$CDM where dark energy is time dependent and $w$ is also varying with time. The latter two models are both models of dark energy which modify the matter content of the Universe. The main method used in H0LICOW papers for measuring distances is first presented in \cite{Para} and later used by Jee, Komatsu and Suyu to measure the angular diameter distances of strong gravitational lenses \cite{Jee}. In this method the angular diameter distance from the observer to the lens were derived by dividing the physical size of the system by the angular separation of lensed image positions. This method while independent of the local distance ladder, is  dependent on the assumption of velocity dispersion and mass profile of the lensing system. It seems worthwhile to conduct the same analysis in various modified gravity models where the geometry of the Universe is no longer necessary Riemannian. Here, we aim to conduct this analysis in Einstein-Cartan cosmology where torsion is present, taking into account that the cosmological distances may be different in a non-Riemann spacetime.

In this paper, we study the cosmological distances and Hubble parameter measurement in the framework of gravity theories with torsion. To illustrate the kind of effects expected in the presence of torsion, we specially focus on the Einstein-Cartan cosmology as the simplest model of this kind. In section \ref{Sec: 2} we review the basic definitions and field equations in Einstein-Cartan cosmology. In section \ref{Sec: 3} by adopting simple ansatzes for the spin density (or torsion) in Einstein-Cartan cosmological equations, we derive relations for measuring various types of cosmological distances and also redshift-time relations in EC cosmology and compare the results to the general relativistic case. Section \ref{Sec: 4} deals with time delay measurement in gravitational lensing systems. Given that angular diameter distances to the lens and the source needed for measuring the Hubble parameter from time-delay distance are different in EC cosmology and General Relativity, we conclude that for certain spin density and torsion ansatz, the measured Hubble parameter will be lower than that of General Relativity.  Section \ref{Sec: 5} is devoted to the discussion of the main results and conclusion. A brief discussion about the same procedure in some typical solutions in the Poincar{\'e} gauge theory of gravity is presented in the appendix.

\section{Einstein-Cartan cosmology}
\label{Sec: 2}
In Einstein-Cartan theory of gravity, the gravitational action is constructed by the curvature scalar, similar to the Einstein-Hilbert action of General Relativity
    \begin{equation}
    \label{eq: action}
    S=\frac{1}{2\kappa}\int d^{4}x\sqrt{-g}g^{\mu\nu}R_{\mu\nu},
    \end{equation}
however here the connection is no longer assumed to be symmetric, with the difference between Levi-Civita connection of General Relativity and the general connection is given by contorsion tensor $K_{\mu\nu}^{~~\alpha}$,
    \begin{equation}
    \label{eq: connection}
    \Gamma^{\alpha}_{\mu\nu}=\tilde{\Gamma}^{\alpha}_{\mu\nu}+K_{\mu\nu}^{~~\alpha},
    \end{equation}
where a tilde denotes the quantity in General Relativity. The torsion tensor, which is defined as the anti-symmetric part of the connection, \textit{i.e.} $T_{\mu\nu}^{~~\alpha}=\Gamma^{\alpha}_{[\mu\nu]}$, is related to the contorsion tensor by the following relation
    \begin{equation}
    K_{\mu\nu\alpha}=T_{\mu\nu\alpha}+T_{\nu\alpha\mu}+T_{\alpha\nu\mu},
    \end{equation}
The field equations are derived by varying the action with respect to the dynamical variables, tetrad and spin connection fields (or equivalently metric and ordinary connection). In Einstein-Cartan gravity, one of the field equations gives an algebraic relation between the torsion tensor and its source, the spin density tensor of the matter.
    \begin{equation}
    \label{eq: torspin}
    T_{\mu\nu\alpha}=-\frac{1}{4}\kappa(2\tau_{\nu\alpha\mu}+g_{\alpha\mu}\tau_{\nu}-g_{\mu\nu}\tau_{\alpha}),
    \end{equation}
where $\tau_{\nu\alpha\mu}$ is the spin tensor of the matter fields and $\tau_{\mu}=\tau^{\nu}_{~~~\mu\nu}$ is called the spin vector. This means that in Einstein-Cartan gravity, torsion is not dynamical \textit{i.e.} does not propagate in vacuum. The other field equation in EC gravity have the same form as in GR
\begin{equation}
\label{eq: FE2}
R_{\mu\nu}-\frac{1}{2}g_{\mu\nu}R=\kappa \sigma_{\mu\nu}-\Lambda g_{\mu\nu}
\end{equation}
where $R_{\mu\nu}$, $R$ and $\sigma_{\mu\nu}$ are the Ricci tensor, Ricci scalar and energy-momentum tensors respectively. The differnce between equation \eqref{eq: FE2} and its General relativistic counterpart is that here $R_{\mu\nu}$ and  $\sigma_{\mu\nu}$ are not necessarily symmetric due to the presence of torsion. In a cosmological background, the large scale homogeneity and isotropy of the Universe forces the torsion and spin tensors to have a specific form in Einstein-Cartan theory. Using the $3+1$ decomposition of the metric \textit{i.e.} $g_{\mu\nu}=h_{\mu\nu}-u_{\mu}u_{\nu}$, where $u_{\mu}$ is timelike 4-velocity and $h_{\mu\nu}$ is the projection tensor, and requiring spatial isotropy and homogeneity, the torsion tensor will acquire the following form \cite{Kranas, Tsamparlis}
    \begin{equation}
    \label{eq: torsion}
    T_{\mu\nu\alpha}=2\phi\,h_{\mu[\nu}u_{\alpha]},
    \end{equation}

This means that in Einstein-Cartan cosmology, the torsion can be characterized by a single time-dependent function $\phi(t)$. Using equation \eqref{eq: torspin}, one can see that the spin tensor $\tau_{\mu\nu\alpha}$ can also be parameterized by a single time-dependent spin density function. In a Friedmann-Robertson-Walker (FRW) type Universe with torsion, the line element is
\begin{equation}
\label{eq: FRW}
 ds^{2}=-dt^{2}+a^{2}(t)\Big[\frac{1}{1-K r^2}dr^2+r^2 d\theta^{2}+r^2 \sin{\theta}^2 d\varphi^{2}\Big]  
\end{equation}

Where $a(t)$ is the scale factor and $K=0,-1,+1$ is the 3-curvature parameter.
Due to the symmetries of a FRW spacetime,  matter energy momentum tensor $\sigma_{\mu\nu}$ in EC cosmology takes the form of a perfect fluid
\begin{equation}
\label{eq: pf}
    \sigma_{\mu\nu}=\rho u_{\mu}u_{\nu}+p g_{\mu\nu},
    \end{equation}
where $\rho$ and $p$ are energy density and isotropic pressure of the perfect fluid respectively.
Following the above discussion and using \eqref{eq: torspin}, \eqref{eq: torsion}, \eqref{eq: FRW} and \eqref{eq: pf} , the Einstein-Cartan field equation \eqref{eq: FE2} leads to the Friedmann-like equations in the form \cite{Kranas}
\begin{equation}
    \label{eq: friedmann1}
    \left(\frac{\dot{a}}{a}\right)^{2}=\frac{1}{3}\kappa \rho-\frac{K}{a^{2}}+\frac{1}{3}\Lambda-4\phi\left(\phi+\left(\frac{\dot{a}}{a}\right) \right),
\end{equation}

\begin{equation}
    \label{eq: friedmann2}
    \frac{\ddot{a}}{a}=-\frac{1}{6} \left(\rho+3p\right) +\frac{1}{3}\Lambda-2 \left(\dot{\phi}+ \left(\frac{\dot{a}}{a}\right)\phi\right),
\end{equation}
where $a(t)$ and $K$ are the scale factor and curvature constant in the Friedmann-Robertson-Walker metric respectively, $\Lambda$ is the cosmological constant and $\rho$ and $p$ are energy density and pressure of the matter content. The above Friedmann equation \eqref{eq: friedmann1} can be transformed to a more useful form by introducing appropriate density parameters
\begin{equation}
\label{eq: densityp}
     \Omega_\Lambda=\frac{\Lambda}{3H^{2}},\;\; \Omega_m=\frac{\kappa \rho_m}{3H^{2}},\,\,\Omega_r=\frac{\kappa \rho_r}{3H^{2}}, \;\;\Omega_K=\frac{-K}{a^{2}H^{2}},
\end{equation}
\begin{equation}
    \Omega_\phi=-4 \left(1+\frac{\phi}{H}\right)\frac{\phi}{H}. \nonumber
\end{equation}

Substituting these density parameters into Friedmann equation \eqref{eq: friedmann1}, we get
\begin{equation}
    \label{eq: friedmann3}
    \Omega_{m}+ \Omega_{r}+ \Omega_{k}+\Omega_{\Lambda}+ \Omega_{\phi}=1.
\end{equation}

The continuity equation which govern the energy conservation will be
    \begin{equation}
    \label{eq: cont}
    \dot{\rho}=-3H\Big(1+\frac{2\phi}{H}\Big)(\rho+p)+\phi(\rho+\kappa^{-1}\Lambda),
    \end{equation}
which reduces to its general relativistic form when $\phi=0$.

\begin{figure}[ht!]
    \label{fig: DcNegative}
    \centering
    \includegraphics[scale=0.4]{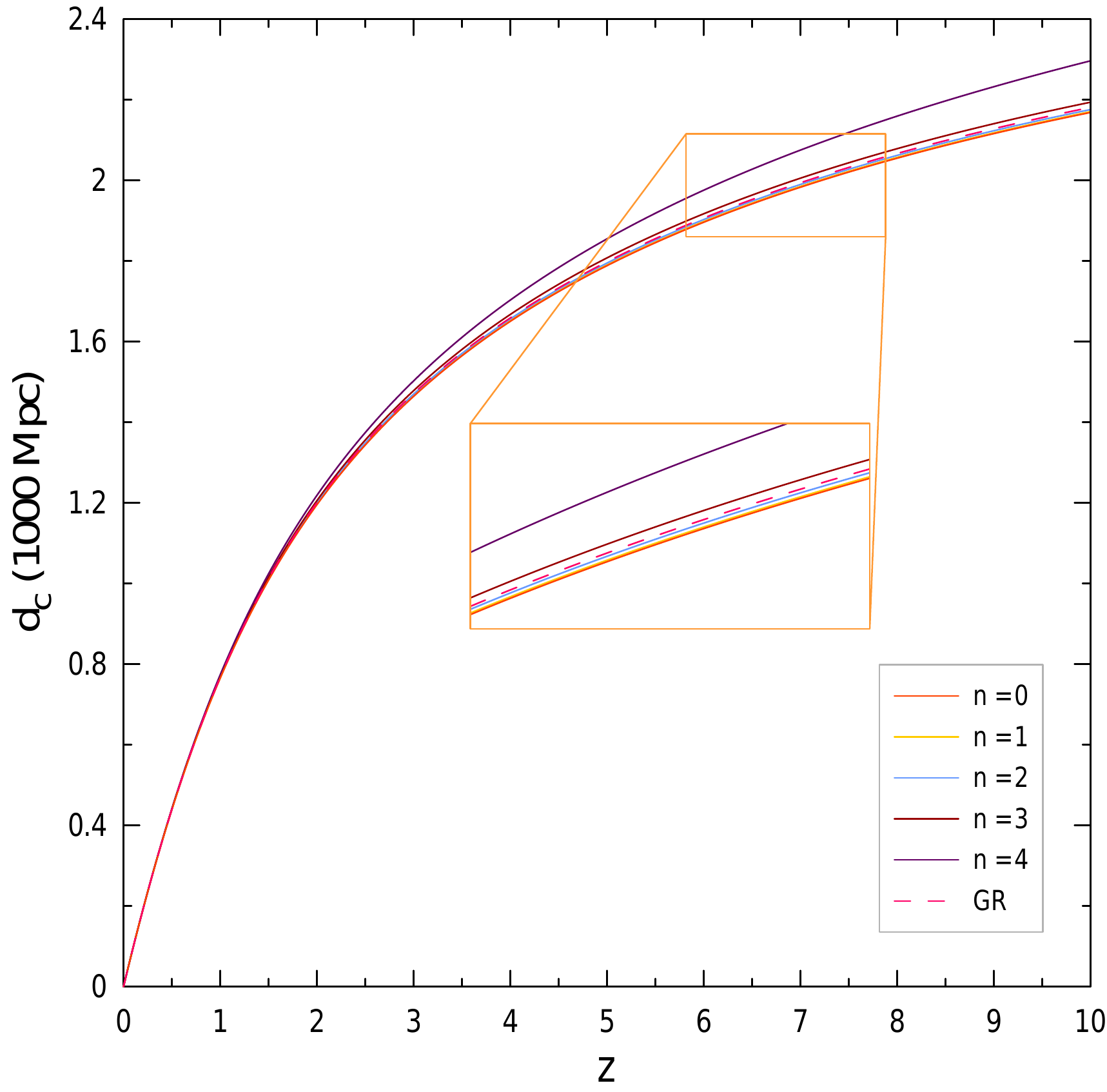}
    \caption{The evolution of the line-of-sight comoving distance versus the redshift for different values of $n$ in equation \eqref{eq: dc}  and negative values of $\Omega_\phi$. }
    \label{fig: DcNegative}
\end{figure}

\begin{figure}[ht!]
    \centering
    \includegraphics[scale=0.4]{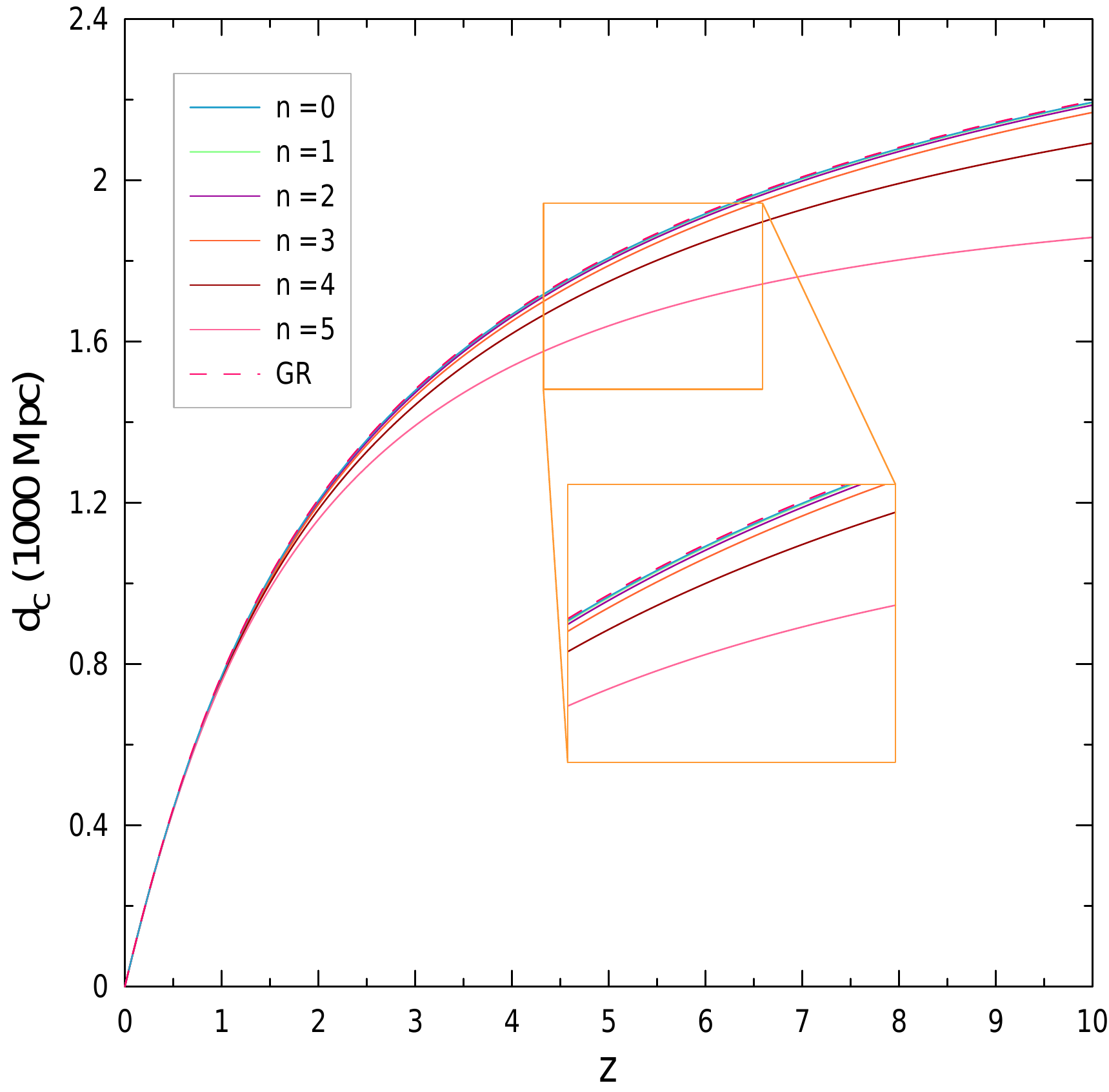}
    \caption{The evolution of the line-of-sight comoving distance versus the redshift for different values of $n$ in equation \eqref{eq: dc} and positive values of $\Omega_\phi$.}
    \label{fig: DcPositive}
\end{figure}
\section{Cosmological Distances in Einstein-Cartan Cosmology}
\label{Sec: 3}
The most fundamental measure of distance in cosmology is called the line-of-sight comoving distance $d_c$. Suppose that we want to measure $d_c$ between a light source and ourselves. By using the Friedmann-Robertson-Walker metric, this will be the proper distance between us and the light source, divided by the ratio of the scale factor at the times of detection and emission of the photons. This obviously depends on the expansion history of the Universe. Generally
    \begin{equation}
    d_c=c\int^{z}_{0} \frac{dz}{H(z)}.
    \end{equation}

In Einstein-Cartan cosmology, using the Friedmann equations above and assuming that torsion density parameter evolves completely independent of matter density parameter, the line-of-sight comoving distance to an object at redshift $z$ is given by
    \begin{equation}
    \label{eq: dc}
    \begin{split}
    d_c=
    &\frac{c}{H_0}
    \int^{z}_{0} {dz
    \over\displaystyle{
    \sqrt{\Omega_m (1+z)^{3}+ \Omega_r (1+z)^{4}+\Omega_\Lambda
    \atop\quad
    {}+ \Omega_K (1+z)^{2}+\Omega_\phi (1+z)^{n}}}},
    \end{split}
    \end{equation}
However, the above assumption is not generally correct as in most available models torsion density depends on the spin of particles and is not independent of matter density. The specific power $n$ in the last term of the square root will depend on the specific model chosen to describe torsion or spin density and will be discussed later. Figure (\ref{fig: DcNegative}) and (\ref{fig: DcPositive}) shows the line-of-sight comoving distance versus the redshift in equation \eqref{eq: dc} for negative and positive values of $\Omega_\phi$ respectively, for various values of $n$. As can be seen from the plots, for negative values of $\Omega_\phi$ and $n=0,1,2$ the line-of-sight comoving distance is less than its GR value for a given redshift, while for $n=3,4$ it is greater than the GR values. For positive values of $\Omega_\phi$ the line-of-sight comoving distance is less than its GR value for a given redshift for all values of parameter $n$.

The line-of-sight comoving distance is closely related to the transverse comoving distance $d_m$ (sometimes called the proper motion distance), which is used to determine the distance between two objects at a same redshift, separated by an angle in the sky. The relation between $d_m$ and $d_c$ is
\begin{align}
    d_m=\frac{c}{H_0}\frac{1}{\sqrt{\left|\Omega_{K}\right|}}\,\, \chi\left(\frac{H_0}{c}\sqrt{\left|\Omega_{K}\right|} d_c\right),\label{2}
\end{align}
where
\begin{align}
    \chi(x)= \begin{cases}\sin (x), & \Omega_{K}<0, \\ x, & \Omega_{K}=0, \\ \sinh (x), & \Omega_{K}>0.\end{cases}\label{3}
\end{align}

Note that the presence of torsion, via the last term in the left-hand-side of the Friedmann equation \eqref{eq: friedmann3} alters the evolution of the scale factor and subsequently changes comoving distances to an object at a given redshift. Comoving distances, while fundamental, can not be measured directly by observations. In Einstein-Cartan cosmology the redshift formula is the same as General Relativity
    \begin{equation}
     \label{eq: redshift}
    1+z=\frac{a(t_0)}{a(t)}=\frac{1}{a(t)},
    \end{equation}
where the scale factor at the present time $t_o$ is normalized to unity. Angular diameter distance, measured using standard 'rulers', is related to the transverse comoving distance via the following relation
    \begin{equation}
    d_A=a(t)\,d_m=\frac{d_m}{1+z}.
    \end{equation}

\begin{figure}[ht!]
    \centering
    \includegraphics[scale=0.4]{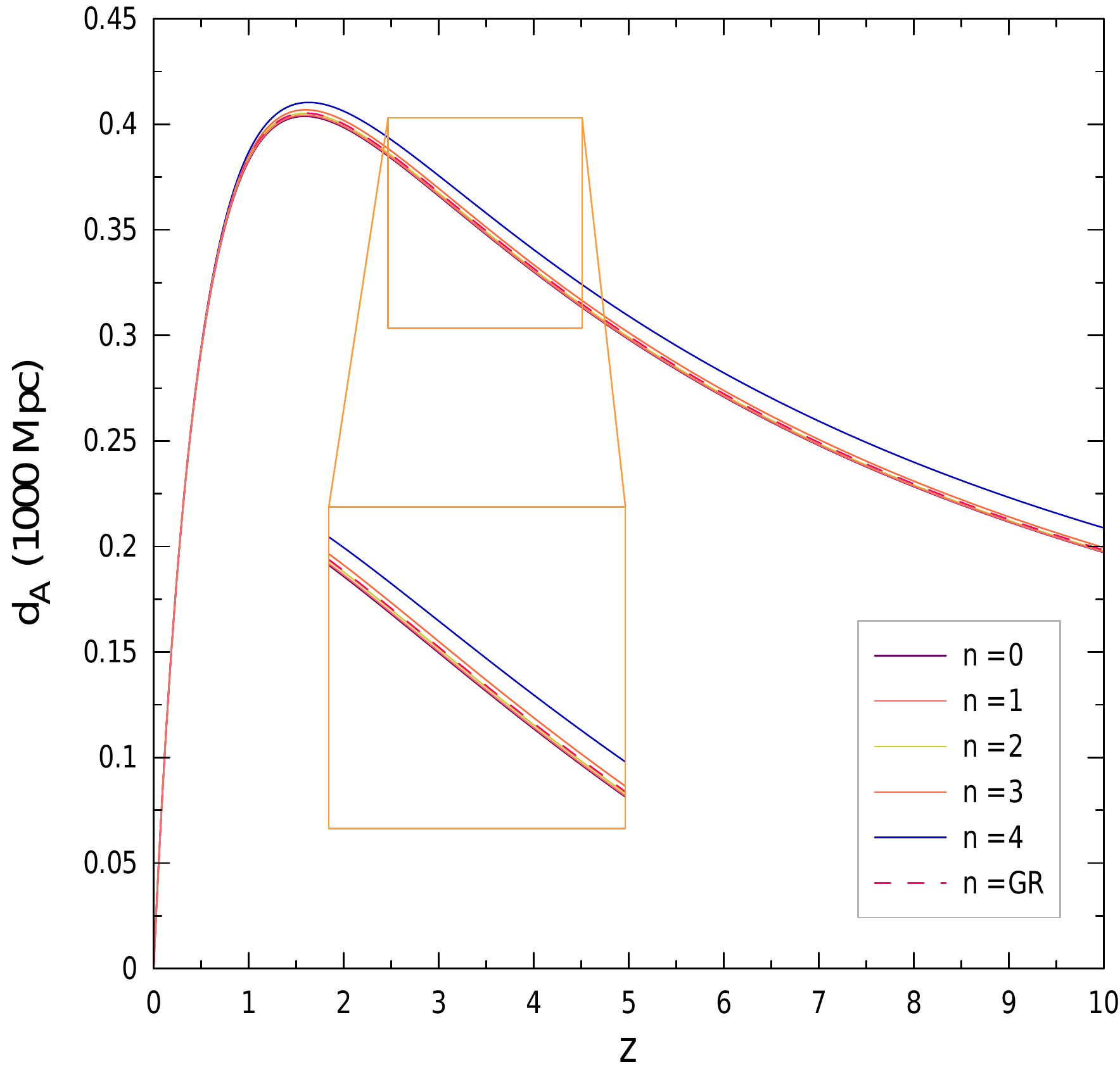}
    \caption{The evolution of the angular diameter distance versus the redshift for different values of $n$ in equation \eqref{eq: dc}  and negative values of $\Omega_\phi$.}
    \label{fig: DaNegative}
\end{figure}

\begin{figure}[ht!]
    \centering
    \includegraphics[scale=0.4]{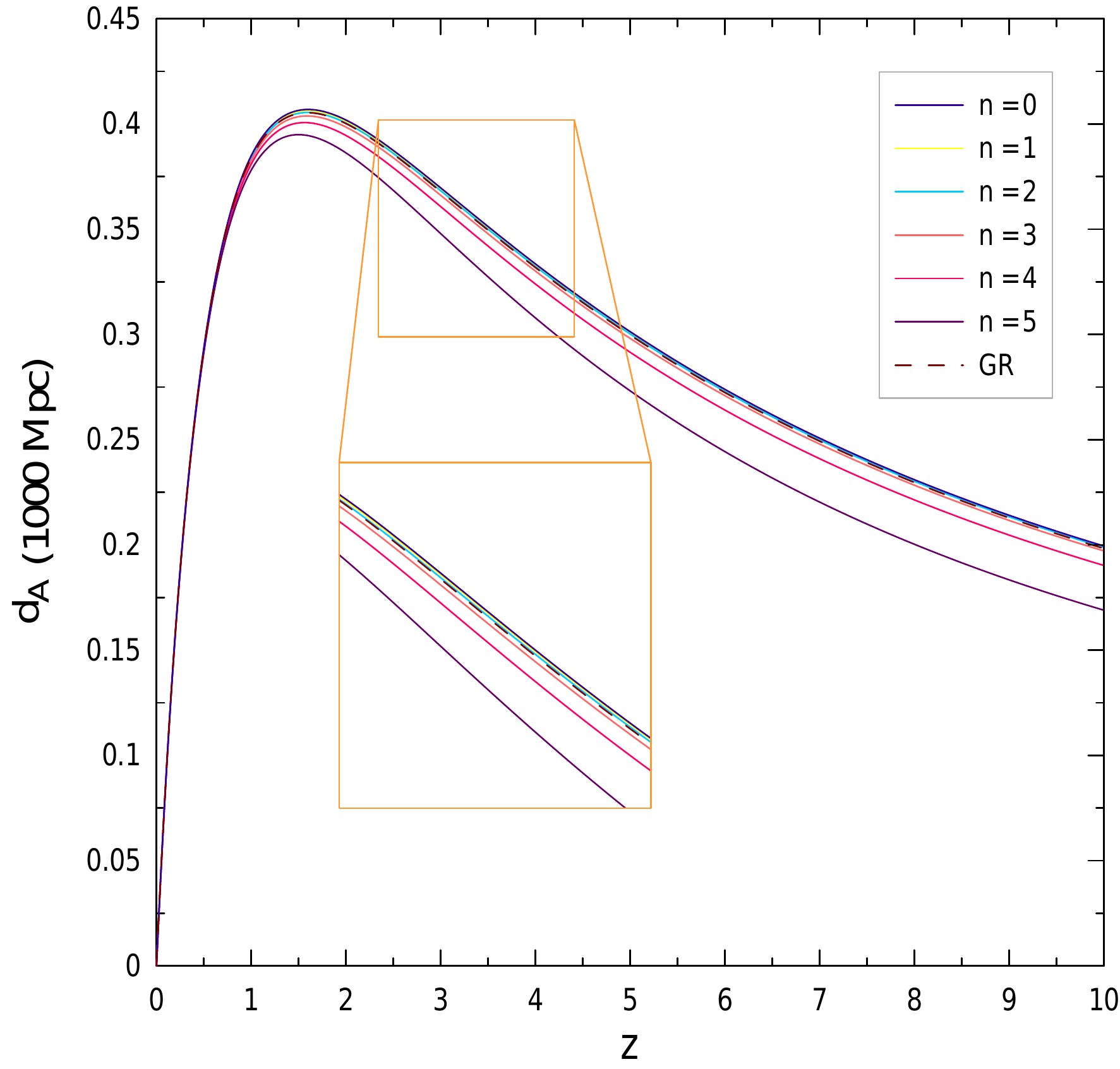}
    \caption{The evolution of the angular diameter distance versus the redshift for different values of $n$ in equation \eqref{eq: dc}  and positive values of $\Omega_\phi$.}
    \label{fig: DaPositive}
\end{figure}

This relation is also the same in EC cosmology and General Relativity \cite{Bolejko}. The angular diameter distance is used to determine distances and time delay in a gravitational lensing system. Due to the effects of the expansion, the angular diameter distance between two objects at different redshifts is not governed by a simple subtraction rule. For example, if we denote the angular diameter distances between the lens and the source, the lens and the observer and the source and the observer in a lensing system by $d_{A(LS)}$, $d_{A(LO)}$ and $d_{A(SO)}$ respectively, then
    \begin{equation}
    \label{eq: dals}
    \begin{split}
    d_{A(LS)}=\frac{1}{1+z_L}&
    \Bigg[d_{m(SO)}\sqrt{1+\frac{\Omega_K\,H^{2}_{0}}{c^{2}}d^{2}_{m(LO)}}
    \\
    &-d_{m(LO)}\sqrt{1+\frac{\Omega_K\,H^{2}_{0}}{c^{2}}d^{2}_{m(SO)}}\Bigg],
    \end{split}
    \end{equation}
where $z_L$ is the redshift of the lens and $d_m$s are respective transverse comoving distances. The above formula is valid for $\Omega_K\geq0$. Figure (\ref{fig: DcNegative}) shows the evolution of the angular diameter distance versus the redshift for different values of parameter $n$ in equation \eqref{eq: dc} and negative values of  $\Omega_\phi$. Figures (\ref{fig: DaNegative}) and (\ref{fig: DaPositive}) show the angular diameter distance in the presence of torsion for negative and positive values of $\Omega_\phi$ respectively. It can be seen from the plots that  for negative values of $\Omega_\phi$ and for a given redshift, the angular diameter distance is greater than its GR value for $n=3,4$ and less than the GR value for $n=0,1,2$ while for positive values of $\Omega_\phi$ it is greater than the GR value for $n=0,1,2$ and less than the GR value for $n=3,4,5$.

Finally, the luminosity distance is defined by using the bolometric luminosity and observed bolometric flux of a source. In practice it is measured by using standard candles, usually  Cepheid Variable stars and Type Ia supernovae. In Einstein-Cartan cosmology, the relationship between luminosity and angular diameter distances is modified due to the presence of torsion \cite{Bolejko}
    \begin{equation}\
    \label{eq: dlda}
    d_l=(1+z)^{2}(1+\eta)\,d_A,
    \end{equation}
where the parameter $\eta$ represents the effects of torsion on the flux and cross section of light rays \cite{Bolejko}. On the limit $\eta\rightarrow 0$ the above relation reduces to that of General Relativity. This relation will impact the calibration of the luminosity distances to Type Ia supernovae using known angular diameter distances to the a lens galaxy, obtained by the method described in \cite{Jee} and used in various estimations of $H_0$, for example in \cite{Wong}.

In Einstein-Cartan cosmology, the exact form of torsion function $\phi$ depends on the specific model chosen to describe the cosmic spin fluid, one of the most used of these being the Weyssenhoff fluid model \cite{Weys}. In any case, generally  the torsion function will depend on the Hubble parameter and scale factor as
    \begin{equation}
    \label{eq: pq}
    \phi=\eta_{0}H^{p}a^{q},
    \end{equation}
where $\eta_0$, $p$ and $q$ are constants. We will now discuss various values of $p$ and $q$ and their effects on cosmological distances and distance-redshift relations.

Differentiating equation \eqref{eq: redshift}
    \begin{equation}
    \label{eq: dzdt}
    dz=\frac{dz}{da}\frac{da}{dt}dt=-\frac{a_0}{a^2}\dot{a}dt=-\frac{1}{H_0}\frac{\dot{a}\dot{a_0}}{a^2}dt,
    \end{equation}

    \begin{itemize}
    \item {Case A: $p=1$, $q=0$}
    \end{itemize}
This case, sometimes called the stead-state torsion, assumes that the contribution of torsion to the volume expansion rate is constant \cite{Kranas}. This for example can happen in a universe with $\rho=K=\Lambda=0$ in the presence of torsion (see Friedmann equation \eqref{eq: friedmann1}). This is not a realistic assumption in a real Universe but as the torsion has the dimension of the Hubble parameter, this can give us a fair idea of what effects we should expect from the existence of torsion. In this case it is obvious from equation \eqref{eq: densityp} that  $\Omega_\phi$ reduces to a constant. From equation \eqref{eq: friedmann1} we have
    \begin{equation}
    \dot{a}^{2}=\frac{\frac{1}{3}\kappa\rho a^{2}-K+\frac{1}{3}\Lambda a^{2}}{1-4\eta_{0}-4\eta_{0}^{2}},
    \end{equation}
which in the case $\eta_{0}=0$ reduces to its general relativistic form. Substituting the above equation in \eqref{eq: dzdt} and using equation \eqref{eq: densityp},  gives us the relationship between $dz$ and $dt$ in this case
\begin{equation}
\begin{split}
dz=&-H_0(1+z)(1-4\eta_0-4\eta^{2}_{0})^{\frac{-1}{2}}\times\Big(\Omega_m (1+z)^3\\
&+\Omega_r (1+z)^4
-\Omega_k (1+z)^2+\Lambda\Big)^{\frac{1}{2}}dt,
\end{split}
\end{equation}
The parameter $\eta$ in equation \eqref{eq: dlda} in this case is \cite{Bolejko}
\begin{equation}
\eta\approx \eta_0 \ln (1+z).
\end{equation}

\begin{figure}[ht!]
    \centering
    \includegraphics[scale=0.4]{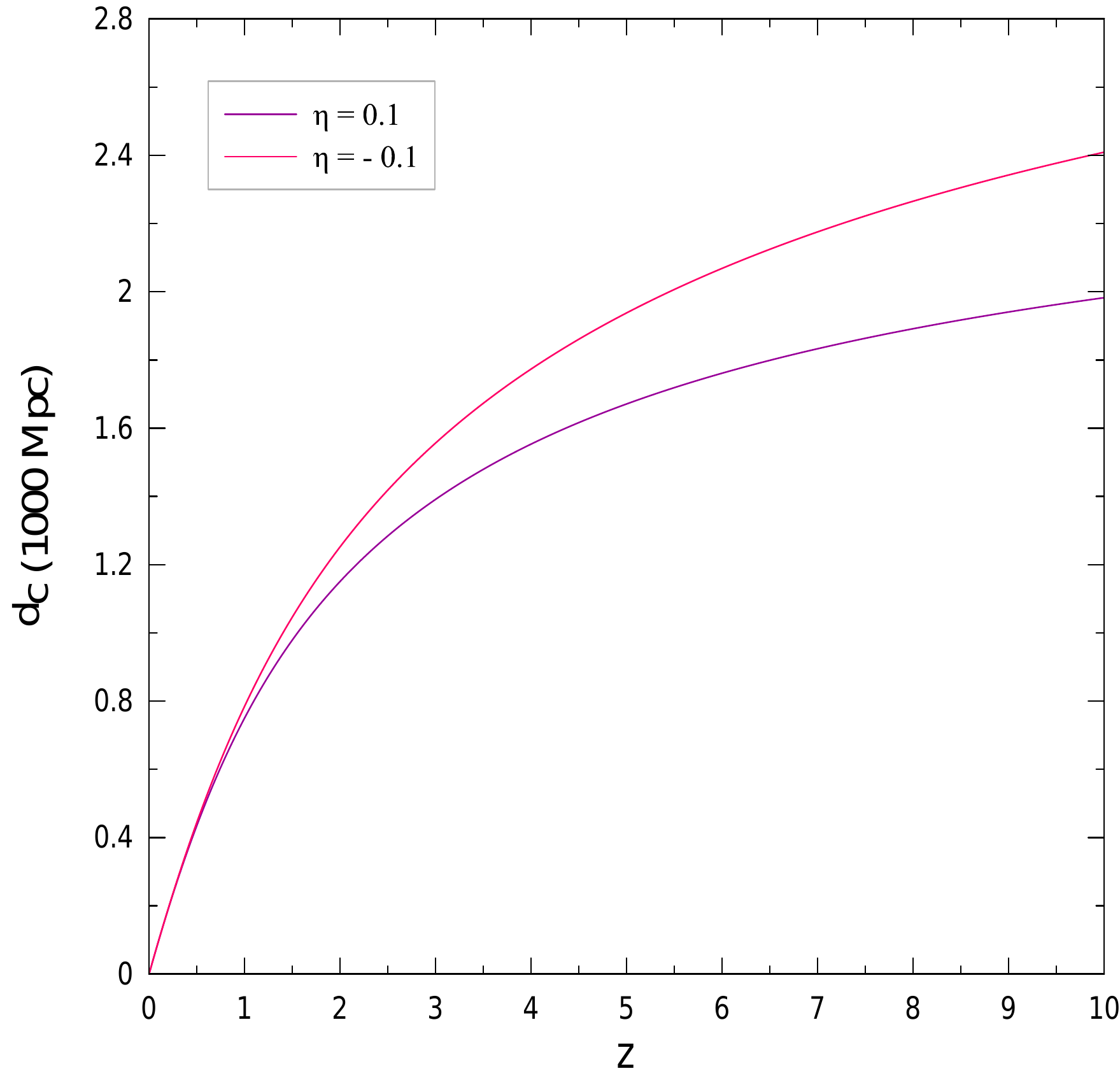}
    \caption{The evolution of line-of-sight comoving distance for two values of parameter  $\eta_0$ in case B in equation \eqref{eq: CaseB}.}
    \label{fig: DcEC}
\end{figure}

\begin{figure}[ht!]
    \centering
    \includegraphics[scale=0.4]{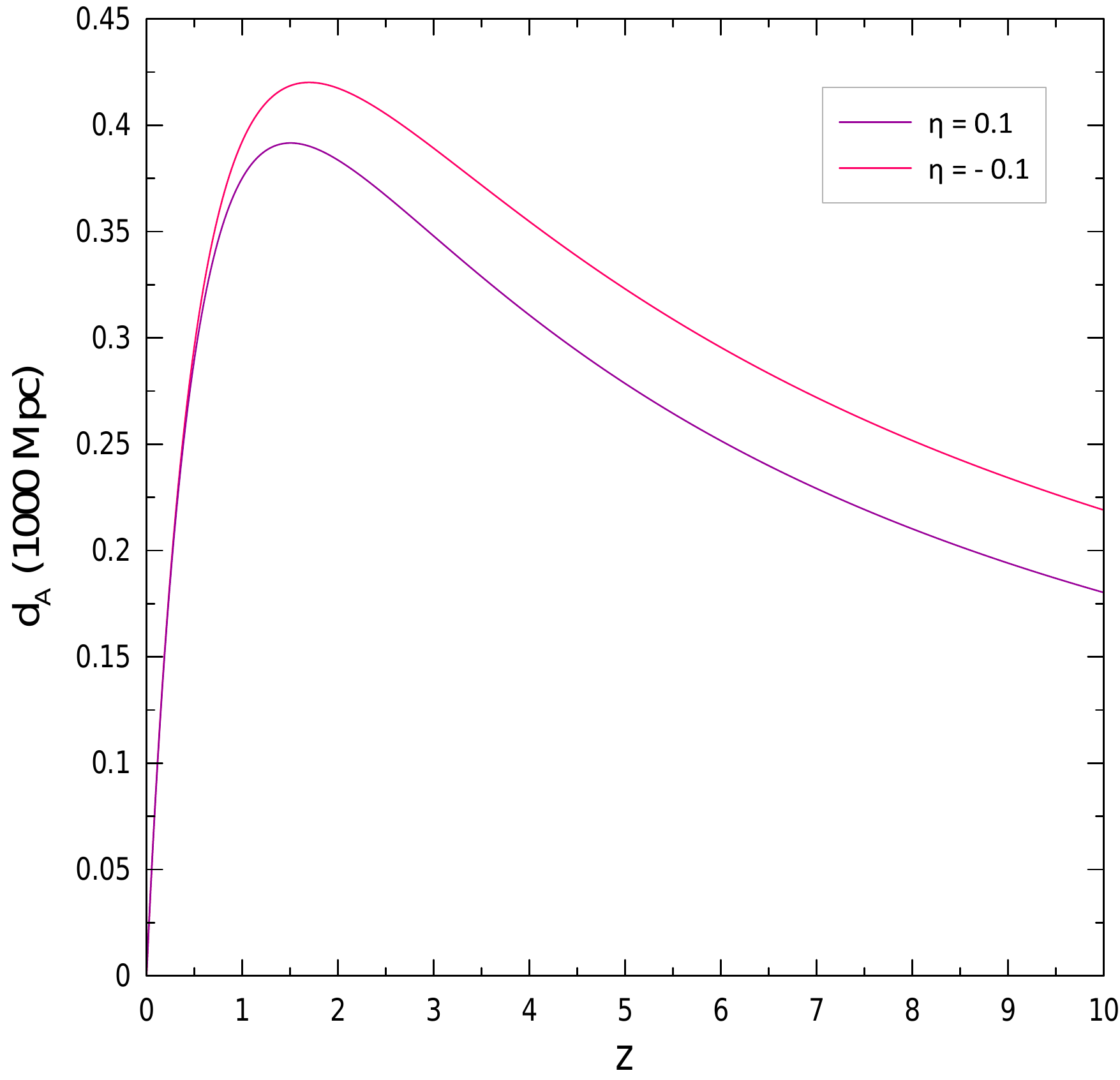}
    \caption{The evolution of angular diameter distance for two values of parameter  $\eta_0$ in case B in equation \eqref{eq: CaseB}.}
    \label{fig: DaEC}
\end{figure}

 \begin{itemize}
    \item {Case B: $p=1$, $q=-1$}
    \end{itemize}
    This is a more realistic generalization of case A which assumes that the contribution of torsion to the dynamics dilutes with the expansion. In this case we have
    \begin{equation}
    \label{eq: CaseB}
    \dot{a}^{2}=\frac{\frac{1}{3}\kappa\rho a^{2}-K+\frac{1}{3}\Lambda a^{2}}{\frac{1}{a^2}-\frac{4\eta_{0}}{a^3}-\frac{4\eta_{0}^{2}}{a^4}}.
    \end{equation}
    The relationship between $dz$ and $dt$ is not as straightforward as the previous case. The comoving and angular diameter distances in this case are given in figures (\ref{fig: DaEC}) and (\ref{fig: DcEC}). The parameter $\eta$ will be
 \begin{equation}
\eta\approx \eta_0 z.
\end{equation}

 \begin{itemize}
    \item {Case C: Weyssenhoff-Raabe spinning fluid model}
     \end{itemize}
 In this case we assume the source of torsion to be unpolarized spin $\frac{1}{2}$ particles \cite{Weys}. The ideal Weyssenhoff fluid is a continuous medium where the fluid "particles" are characterised in addition to their energy and momentum, by their intrinsic angular momentum (spin) proportional to the volume. The spin density is described by the second-rank anti-symmetric tensor
\begin{equation}
\tau^{\mu\nu}=-\tau^{\nu\mu}
\end{equation}
This tensor also obeys what is called the Frenkel condition
 \begin{equation}
 \tau^{\mu\nu}u_{\nu}=0
 \end{equation}  
  where  $u_{\nu}$ is the 4-velocity.  In the EC theory this spin takes on an important role as the source of the torsion part of the gravitational field.
  Using the variational formalism first introduced by  Ray and Smalley \cite{Ray1, Ray2, Ray3, Ray4, Ray5, Smalley1},  the canonical spin tensor in EC theory is given by
  \begin{equation}
  \tau^{\mu\nu\rho}=\frac{1}{2}\tau^{\mu\nu}u^{\rho}
  \end{equation}
  where $\tau^{\mu\nu}$ and $u^{\rho}$ are spin density and 4-velocity of the fluid respectively. In this way the dynamical energy-momentum can be decomposed into the usual perfect fluid part and an intrinsic spin part.  If we associate the spin density tensor in EC cosmology to the quantum-mechanical spin of microscopic particles, an appropriate space-time averaging of the dynamical energy-momentum tensor is needed to define the effective sources of the macroscopic gravitational interactions.  One should note that even in the case of randomly oriented spins of consisting particles, the average of the spin-squared terms generally will not be zero. For this reason, the Einstein-cartan field equations are different from their general relativistic counterparts, even in the classical macroscopic limit. Using the averaging procedure discussed in \cite{Gas, Nur}, the square of the spin density $\tau^{2}=\frac{1}{2}\langle \tau_{\mu\nu\rho}\,\tau^{\mu\nu\rho}\rangle$ will depend on the matter energy density $\rho$ by the relation

 \begin{equation}
 \tau^{2}=\frac{1}{8}\hbar^{2}A_{\omega}^{-2/1+\omega}\rho^{2/1+\omega}
 \end{equation}
 where $A_{\omega}$ is a constant depending on the equation of state parameter $\omega$. If we assume that the spin of the fermions in the spinning fluid are not polarized (which will not be the case if there exist primordial magnetic fields in the early universe), then the relation between the energy density $\rho$ and the scale factor will be the same as in GR i.e. $\rho \propto a^{-3(1+\omega)}$. With this assumptions, the relation between torsion function $\phi$ and the scale factor will be
 \begin{equation}
 \phi^{2}\propto a^{-6}
 \end{equation}
 which is independent of the equation-of-state parameter $\omega$ but note that $A_{\omega}$ in the proportionality constant still depends on $\omega$. Comparing with \eqref{eq: pq}, we can see that for the Weyssenhoff-Raabe spinning fluid model $p=0$ and $q=-6$. From the above discussion and the relation \eqref{eq: densityp}, one can see that in this model, torsion density parameter term in the integral  \eqref{eq: dc} consists of two terms, depending on the redshift as $(1+z)^3$ and $(1+z)^6$.  The first term can be absorbed into the matter density term and for the second term we have $n=6$ in \eqref{eq: dc}. It should be noted that this is a semi-classical model and should not be regarded as the complete description of a spinning fluid as it does not consider other fundamental interactions besides gravity. Nonetheless, it is a widely used and very useful model to analyze the effects of torsion on cosmological dynamics.

\section{Hubble parameter from gravitational lensing time delay}
\label{Sec: 4}
 In  a gravitational lensing system, the arrival time of light rays coming from two different images will be different if they pass through regions with different gravitational potential. This effect, called the Shapiro time delay can be used to determine the Hubble parameter from suitable lensing systems \cite{Refsdal}. In a universe described by Friedmann-Robertson-Walker metric, the time delay between two images is given by \cite{Schneider}
\begin{equation}
\label{eq: td}
    \Delta t_{12}=\frac{(1+z_L)D_{\Delta t}}{c} \Delta \Phi_{12},
\end{equation}
where $z_L$ is the redshift of the lens, $\Delta \Phi_{12}$ is the difference in the Fermat potential between the two images and $D_{\Delta t}$ is the time-delay distance given by
    \begin{align}
    \label{eq: tdd}
    D_{\Delta t} \equiv\left(1+z_{L}\right) \frac{d_{A(LO)}\, d_{A(SO)}}{d_{A(LS)}}=\frac{c}{H_{0}} \frac{\bar{d}_{m(LO)}\, \bar{d}_{m(SO)}}{\bar{d}_{m(LS)}},
    \end{align}
 where $\bar{d}_{m(LO)}$, $\bar{d}_{m(SO)}$ and $\bar{d}_{m(LS)}$ are dimensionless transverse comoving distances (\textit{i.e.} comoving distances divided by the Huble distance $c/H_0$). Usually, the measured distances in a lensing system are luminosity distances obtained with the help of standard candles. If Luminosity distances to the lens and the source are known, we can use equation \eqref{eq: dlda} to determine angular diameter distances
\begin{equation}
    d_{A(LO)}=\frac{d_{l(LO)}}{(1+z_{L})^2 \: (1+\eta_{L})},
\end{equation}
\begin{equation}
    d_{A(SO)}=\frac{d_{l(SO)}}{(1+z_{S})^2 \: (1+\eta_{S})},
\end{equation}
and then $ d_{A(LS)}$ is derived using equation \eqref{eq: dals}.

Using the above equations, the time delay distance in equation \eqref{eq: tdd} will be
\begin{equation}
    D_{\Delta t} =\frac{
    (1+z_{L}) d_{l(SO)} d_{l(LO)}}{(1+z_{L})^2 (1+z_{S})^2
    (1+\eta_{S})
    (1+\eta_{L})
    \bar{d}_{LS}},
\end{equation}
and time delay in equation \eqref{eq: td} is
\begin{equation}
\label{eq: td0}
    \Delta{t}= \frac{d_{l(SO)} d_{l(LO)} \Delta{\Phi}}{c (1+z_S)^2 (1+\eta_{S}) (1+\eta_{L})d_{A(LS)}}.
\end{equation}

If the luminosity distances and redshifts of the lens and the source are known, one can use the above equation to determine the Hubble parameter and the curvature density parameter from the time delay between the images. Note that the torsion also effects the Fermat potential in equation \eqref{eq: td} as it can alter the position of the images and the lens potential. These types of effects have been studied previously, for example in Reference \cite{Zamani} for a particular Schwarzschild lens, however, the impact of these modifications to the Fermat potential on the measurement of the Hubble parameter is negligible compared to the effects of torsion on cosmological distances.

For the case A above we have for the torsion function $\phi$
\begin{equation}
    \phi=\eta_{0} H_{0} \: \rightarrow \: \eta \thickapprox \eta_{0}z,
\end{equation}
substituting this in equation \eqref{eq: td0}
\begin{equation}
\label{eq: td1}
\Delta{t}=  \frac{{d_{l(SO)}\, d_{l(LO)}\,\Delta\Phi}}{c \: (1+z_S)^2 \: (1+\eta_{0} z_{S}) \: (1+\eta_{0} z_{L})\: d_{A(LS)}}.
\end{equation}
Similarly for the Case B we have
\begin{equation}
    \phi=\eta_{0} H \: \rightarrow \: \eta \thickapprox \eta_{0} \ln(1+z),
\end{equation}
so
\begin{equation}
\label{eq: td2}
\begin{split}
\Delta{t}&=\frac{{d_{l(SO)}\, d_{l(LO)}\,\Delta\Phi}}{c \: (1+z_S)^2 \:(1+\eta_{0} (\ln{(1+z_{S})}))} \\
 &\times \frac{1}{(1+\eta_{0}) (\ln{(1+z_{L})})d_{A(LS)}}
\end{split}
\end{equation}
Equations \eqref{eq: td1} and \eqref{eq: td2} can be used to determine the Hubble parameter in these two cases if the redshifts and distances of the source and the lens are known. For the Weyssenhoff-Raabe model (Case C), we could also directly use integral \eqref{eq: dc} to determine the cosmological distances and measure the Hubble parameter.

In our subsequent calculations of the Hubble parameter from the lensing time delays, we adopt a flat geometry with $K=\Omega_K=0$. Table 1 shows the value of time delays between different images, redshift of the lens, and redshift of the source for seven well-known gravitational lensing systems. In order to compare our results to previous works done in general relativity, for each individual lensing system in table 1, we adopted the exact lensing configuration and lensing potentials specified in references mentioned in the table. The first system, DES J0408-5354 is a strongly lensed quasar where the lens is a red galaxy. This system produces two sets of multiple images at different redshifts. The value of $H_0$ for this system in General Relativity, assuming a flat $\Lambda$CDM model is measured in \cite{Shajib} as  $74.2^{+2.7}_{-3}$ $km \,s^{-1}\,Mpc^{-1}$. Here, assuming EC cosmology and choosing the ansatz for the torsion as in case B above, we evaluate the modified angular diameter distances and use equation  \eqref{eq: td2}  to determine the value of $H_0$ as $71.6^{+2.5}_{-2.8}$ $km \,s^{-1}\,Mpc^{-1}$, lower than its GR value (the parameter $\eta_0$ in this and all subsequent calculations is set to $-0.1$). The second system,  WFI2033-4723 is a quadruply-imaged gravitationally lensed quasar where the lens is a massive elliptical galaxy. The GR value of $H_0$ again assuming a flat $\Lambda$CDM cosmology is measured in \cite{Rusu} as $71.6^{+3.8}_{-4.9}$ $km \,s^{-1}\,Mpc^{-1}$ while using the same procedure described for the previous system we arrive at  $H_0=69.2^{+3.6}_{-4.6}$ $km \,s^{-1}\,Mpc^{-1}$ in Einstein-Cartan theory. The next system, SN Refsdal, is a supernova gravitationally lensed by the galaxy cluster MACS J1149.6+2223 with multiple images. The GR value of $H_0$ measured in \cite{Grillo} is $73.5\pm 4.3$ $km \,s^{-1}\,Mpc^{-1}$ while its value in EC cosmology is $71.5\pm 4.1$ $km \,s^{-1}\,Mpc^{-1}$. The system HE0435-1223 is also a quadruply-lensed quasar where the main lens is a galaxy residing in a group of at least 12 galaxies. The GR value of  $H_0$ measured in \cite{Bonvin} is $71.9^{+2.4}_{-3}$ $km \,s^{-1}\,Mpc^{-1}$ (with slightly lower value given in \cite{Chen}) while in EC cosmology we get $H_0=69.1^{+2.3}_{-2.9}$. The next source, and also quadruply-imaged quasar RXJ1131-1231 is lensed by a galaxy and its GR value of $H_0$ is measured in \cite{Chen} as $77^{+4}_{-4.6}$ $km \,s^{-1}\,Mpc^{-1}$ while in EC cosmology we get  $H_0=74.8\pm 2.2$. The system PG1115+080 again a quadruply-imaged quasar lensed by a galaxy gives the GR value of $H_0$ as $82.8^{+9.4}_{-8.3}$ $km \,s^{-1}\,Mpc^{-1}$ \cite{Chen} while its EC value is derived here as  $78.9\pm 9$ $km \,s^{-1}\,Mpc^{-1}$. The last system studied here, MRG-M0138  is likely a Type Ia supernova gravitationally lensed by a foreground galaxy cluster. The measured value of $H_0$ for this system in EC cosmology is measured here as $69.8^{+2.5}_{-2.7}$ $km \,s^{-1}\,Mpc^{-1}$. As can be seen from  all of the studied lensing systems, for Case B with negative values  of torsion parameter , the value of the $H_0$ will be lower in Einstein-Cartan cosmology compared to its general relativistic measurements. The values of Hubble parameter for the seven systems mentioned above using the Weyssenhoff-Raabe model (Case C) are also given in the table. As can be seen from the table, for negative torsion the values of $H_0$ in this case are also lower than their GR values but by a lesser amount compared to case B.

Table 2 shows the constraints on the present day Hubble parameter $H_0$, Matter energy density $\Omega_m$ and the dark energy density $\Omega_\Lambda$ from the combined time delay cosmography data of seven individual lenses in table 1. For this analysis we adopted a uniform prior for $H_0$ in the range $[0,150]$ $km \,s^{-1}\,Mpc^{-1}$, $\omega_m$ in the range $[0.05,1]$, and $\omega_{\phi}$ in the range $[-0.05,0.05]$ with $\Omega_\Lambda=1-\Omega_m-\Omega_{\phi}>0$. As can be seen from the table, while time delay cosmography provides a strong constraint on $H_0$, the estimation of density parameters have large amount of uncertainties. This is the result of the fact that the time delay distance $D_{\Delta t}$ is only weakly sensitive to the density parameters. The median values obtained for GR, EC (Case B) and EC (Case C) are $73.6^{+1.6}_{-1.5}$, $70.8^{+1.4}_{-1.3}$ and $72^{+1.5}_{-1.5}$ $km \,s^{-1}\,Mpc^{-1}$ respectively. As the value of the torsion density parameter $\Omega_{\phi}$ is expected to be very low in late universe,  it can not be meaningfully constrained by time delay cosmography.
\begin{table*}[h]
\label{table:time_delays}
\begin{tabular*}{\textwidth}{@{\extracolsep\fill}lcccccc}
\toprule%
Source name & Time delay (days) & $z_L$ & $z_S$ & $H_0$ (GR) & $H_0$ (With torsion) \\
  &  &  &  & $km(s\,Mpc)^{-1}$ & $km(s\,Mpc)^{-1}$ \\
\midrule
 DES J0408-5354 & $\Delta{t}_{AB}=-112.1 \pm 2.1 $ & 0.597 & 2.375 & $74.2^{+2.7}_{-3}$ &(B)  $71.6^{+2.5}_{-2.8}$\\
\cite{Shajib, Lin, Courbin}&  $\Delta{t}_{AD}=-155.5 \pm 12.8$ & & & &(C)  $73.3^{+2.5}_{-3}$\\
 & $\Delta{t}_{BD}=-42.4 \pm 17.6$ &  &  \\

 \\
 WFI2033-4723 & $\Delta{t}_{B-A_{1}}=-36.2^{+1.6}_{-2.3}$ & $0.6575$ &
1.662 &$71.6^{+3.8}_{-4.9}$&(B)  $69.2^{+3.6}_{-4.6}$\\
\cite{Rusu} & $\Delta{t}_{B-A_{2}}=-37.3^{+2.6}_{-3.0}$ & & & & (C)  $70.8^{+3.7}_{-4.8}$ \\
 & $\Delta{t}_{B-c}=-59.4 \pm 1.3$ & & & \\

 \\
SN 'Refsdal &$\Delta{t}_{S2:S1}=4\pm 4$ & 0.54 & $1.488 $  &$73.5\pm 4.3$&(B)  $71.5\pm 4.1$ \\
   \cite{Rodney, Grillo} &$\Delta{t}_{S3:S1}=2\pm 4$& & & & (C)  $72.6\pm 4.2$\\
  &$\Delta{t}_{S4:S1}=24\pm 5$&\\

 \\
HE0435-1223 & $\Delta{t}_{AB}= -8.8\pm 0.8$ & 0.3098 & 1.722 &  $71.9^{+2.4}_{-3}$ &(B) $69.1^{+2.3}_{-2.9}$\\
\cite{Bonvin, Chen}&$\Delta{t}_{AC}= -1.1\pm 0.7$& & & &  (C) $70.5^{+2.3}_{-3}$\\
&$\Delta{t}_{AD}= -13.8\pm 0.9$ & & &\\

\\
RXJ1131-1231 & $\Delta{t}_{AB}= 0.7\pm 1.2$ & 0.295 & 0.657 & $77^{+4}_{-4.6}$&(B) $74.8\pm 2.2$  \\
\cite{Chen, Tewes}&$\Delta{t}_{AC}= 1.1\pm 1.5$& & & &  (C) $75.9\pm 3.1$ \\
&$\Delta{t}_{AD}= -90.6\pm 1.4$ & & &\\
\\
PG1115+080 & $\Delta{t}_{AB}= 8.3^{+1.5}_{-1.6}$ & 0.3098 &1.722 &$82.8^{+9.4}_{-8.3}$ &(B) $78.9\pm 9$  \\
\cite{Bonvin2, Chen}&$\Delta{t}_{AC}= 9.9^{+1.1}_{-1.1}$& & & &(C) $80.4\pm 9.1$\\
&$\Delta{t}_{AD}= 18.8^{+1.6}_{-1.6}$ & & &\\
\\
MRG-M0138 & $\Delta{t}=114^{+28}_{-31}$ & 0.338 & 1.95& $-$ & $(B)  69.8^{+2.5}_{-2.7}$ \\
 \cite{Rodney2}& $\Delta{t}=-17^{+16}_{-19}$ & & & &(C)  $70.9^{+2.7}_{-2.8}$\\
\botrule
\end{tabular*}
\caption{Hubble parameter measurement for some well-studied gravitational lensing systems. The first column gives the name of the source and references where the table values are given. The next three columns give the measured time delays, redshift of the lens, and redshift of the source respectively, as given in the mentioned references. The fifth column gives the measured value of $H_0$ in the framework of $\Lambda$CDM cosmology and General Relativity, as measured in the aforementioned references. The last column gives the results of this paper where the effects of torsion in the measurement of cosmological distances are considered in determining the value of $H_0$. The $H_0$ value in the last column is derived for the case B mentioned in equation \eqref{eq: td2} and value of $\eta_0$ is assumed to be $-0.1$ and Case C with $A_{\omega}=-0.1$.}
\end{table*}

%%%%%%%%%%%%%%%%%%%%%%%%%%%%%%%%%%%%%%%%%

%%%%%%%%%%%%%%%%%%%%%%%%%%%%%%%%%%%%%%%%%
\begin{table*}[h]
\label{table:median}
\begin{tabular*}{\textwidth}{@{\extracolsep\fill}lcccc}
\toprule%
Gravity model & $H_0$ [$km(s\,Mpc)^{-1}$] & $\Omega_m$ & $\Omega_{\Lambda}$  \\
\midrule
General Relativity & $73.6^{+1.6}_{-1.5}$ & $0.30^{+0.14}_{-0.14}$ & $0.70^{+0.16}_{-0.15}$ \\
Eistein-Cartan Case B & $70.8^{+1.4}_{-1.3}$ & $0.31^{+0.15}_{-0.15}$ & $0.69^{+0.15}_{-0.15}$ \\
Eistein-Cartan Case C & $72^{+1.5}_{-1.5}$ & $0.31^{+0.15}_{-0.15}$ & $0.69^{+0.15}_{-0.15}$ \\
\botrule
\end{tabular*}
\caption{Constraints on $H_0$, $\Omega_m$ and $\Omega_\Lambda$ derived from combined data from the seven lenses in table 1. The reported values are medians. In all models we assume $K=\Omega_K=0$.}
\end{table*}

%%%%%%%%%%%%%%%%%%%%%%%%%%%%%%%%%%%%%%%

\section{Conclusions}
\label{Sec: 5}
The accurate measurement of the present day Hubble parameter $H_0$, from both early universe cosmic microwave background data and late universe methods like gravitational lensing time delays is highly dependent on the availability of precise cosmological distances. For CMB data the shift parameter and multipole position of the first CMB peak is reliant on the angular diameter distance to recombination \cite{Barker}. Moreover, in order to measure $H_0$ using gravitational lensing time delays, angular diameter distances to the lens and the source and between the lens and the source should be known. All of these angular diameter distances depend on the expansion history of the universe and are dependent on the underlying cosmological model. Most of the existing research on the estimation of $H_0$ from gravitational lensing (For example the works of SH0ES and H0LICOW teams) primarily focus on $\Lambda$CDM and various dark energy models. It would be interesting to perform the same analysis in various modified gravity theories and compare the result to general relativity.  Here we study the Einstein-Cartan gravity, the simplest gravity theory with torsion. We show that with the presence of torsion, both line-of-sight comoving distance $d_c$ and angular diameter distance $d_A$ to an object in a given redshift can be different from their GR values. In both Einstein-Cartan gravity and Poincar{\'e} gauge theory, certain ansatzes for torsion can lead to values of $d_A$ to a source or a lens in given redshifts to be  greater than their GR value. This leads to the value of $H_0$ slightly less than the value expected in a flat $\Lambda$CDM model in general relativity and can help to ease the Hubble tension. In the presence of torsion, in addition to the distances themselves, the relation between luminosity and angular diameter distances is also modified.  Also, as argued in \cite{Barker} from an early universe perspective, the torsion can act as a dark radiation component and alter the early expansion rate. This helps to bring the value of $H_0$ from CMB data closer to its value obtained by late time measurements, further alleviating the tension problem. The accurate measurement of $H_0$ from lensing systems requires precise observational data on time delays and angular diameter distances as well as the mass profile and velocity dispersion of the lens (usually acquired by high resolution spectroscopy of the lensing galaxy). The precise measurement of the time delay distance also requires analysis of the mass distribution along the line-of-sight to the lens in order to mitigate the effects of external convergences on the time delay distance. In theory, most of these parameters can be modified by the presence of torsion. In particular the mass modelling and potential of the lens can be different when the torsion is present. Studying these effects in gravity theories with torsion can help in understanding and potentially solving the Hubble tension problem. 
%%%%%%%%%%%%%%%%%%%%%%%%%%%%%
\appendix
\section{Comsological distances and Hubble parameter measurement in Poincar{\'e} gauge theory of gravity}
\begin{figure}[h!]
    \centering
    \includegraphics[scale=0.4]{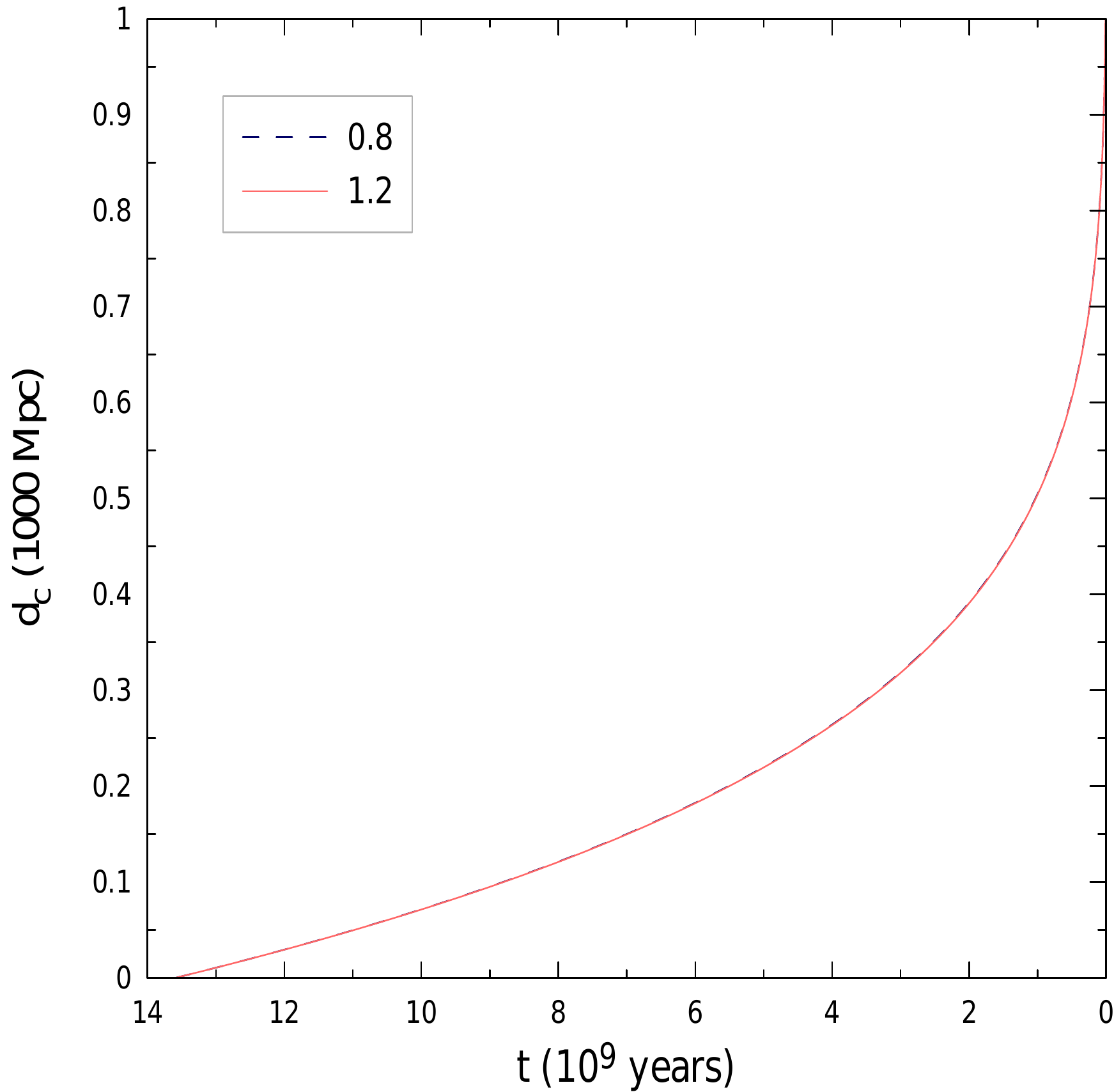}
    \caption{The evolution of line-of-sight comoving distance versus the cosmic time in Poincar{\'e} gauge theory of gravity for the scale factor given by equation \eqref{eq: PGT1sf}. The present-day time is set to $t_0=0$ in this plot and the time axis unit is in Giga years so the horizontal axis shows the age of the universe.}
    \label{fig: PGT1Dc}
\end{figure}

\begin{figure}[h!]
    \centering
    \includegraphics[scale=0.4]{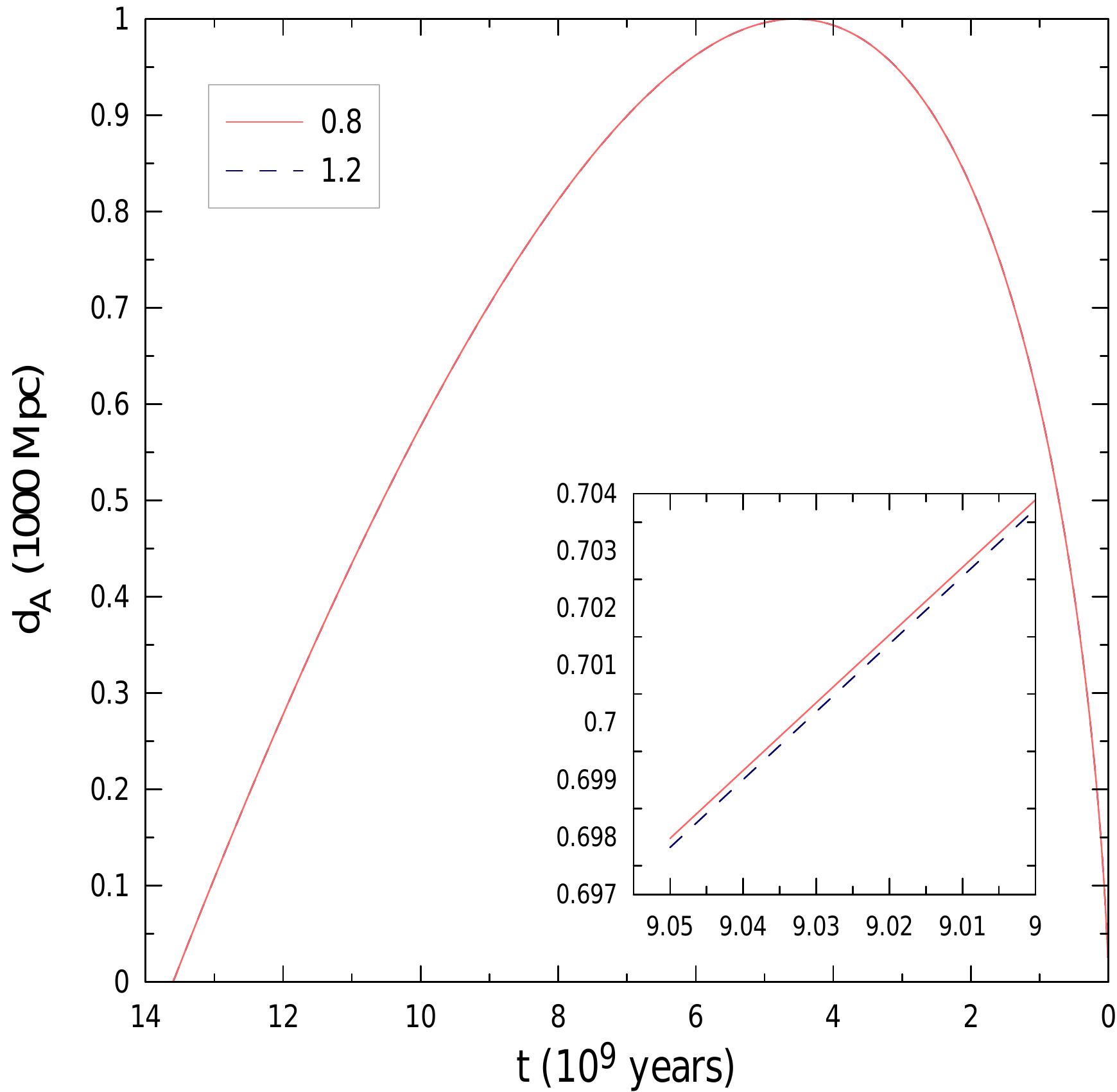}
    \caption{The evolution of angular diameter distance versus the cosmic time in Poincar{\'e} gauge theory of gravity for the scale factor given by equation \eqref{eq: PGT1sf}. The present day time is set to $t_0=0$ in this plot and the time axis unit is in Giga Years so the horizontal axis show the age of the universe.}
    \label{fig: PGT1Da}
\end{figure}

%\input{chapters/appendix}
%\label{Sec: 5}
Poincar{\'e} gauge theory of gravity is a logical extension of both General Relativity and Einstein-Cartan theory where the most general gravitational Lagrangian (with even parity terms) is given by
\begin{equation}
L_g=L_R+L_{R^2}+L_{T^2},
\end{equation}
where
\begin{equation}
L_R=-\frac{1}{2}a_0\,R,
\end{equation}
is the usual Einstein-Hilbert Lagrangian in the presence of torsion and
\begin{equation}
\label{eq: Lagr}
\begin{split}
L_{R^2}=& a_{1}R^{2}+a_{2}R_{\mu\nu}R^{\mu\nu}+a_{3}R_{\mu\nu}R^{\nu\mu}
\\
&+a_{4}R_{\alpha\nu\rho}R^{\alpha\mu\rho\nu}
+a_{5}R_{\alpha\mu\nu\rho}R^{\alpha\nu\mu\rho}
\\
&+a_{6}R_{\alpha\mu\nu\rho}R^{\nu\rho\alpha\mu},
\end{split}
\end{equation}
\begin{equation}
\label{eq: Lagt}
L_{T^2}=b_{1}T_{\alpha\mu\nu}T^{\alpha\mu\nu}+b_{2}T_{\alpha\mu\nu}T^{\mu\alpha\nu}+b_{3}T_{\mu}T^{\mu},
\end{equation}
are quadratic Riemann and torsion sectors respectively. The ten dimensionless parameters of the theory, $a_{0}...a_{6}$ and $b_{1}...b{3}$ can be constrained by various physical considerations such as the requirement that there should be no ghosts and no tachyons in the theory. For a great analysis of the various cases see reference \cite{Barker}. Here we are interested in the measurement of cosmological distances and its implications for the Hubble tension problem in some typical cosmological solutions in Poincar{\'e} gauge theory of gravity. In cosmological applications, the torsion tensor which obeys the symmetries of an FRW spacetime is usually assumed to have the form \cite{Barker, Tsam}
\begin{equation}
\label{eq: Torpg}
T^{~~\alpha}_{\mu\nu}= \frac{1}{3}(\delta^{\alpha}_{\nu}u_{\mu}-\delta^{\alpha}_{\mu}u_{\nu})\,\Phi+\epsilon_{\mu\nu}^{~~\alpha\rho}u_{\rho}\Psi,
\end{equation}
where $\Phi$ and $\Psi$ are torsion functions that can only depend on time due to spacetime symmetries. We examine two recent solutions in viable PGT cosmologies presented in \cite{Barker}. The first called Class $^{3}C^{\ast}$ in \cite{Barker} have a general solution for the scale factor in the form
%%%%%%%%%%%%%%%%%%%%%%%%%%%%%%%%%%%%%%%%%
\begin{equation}
\begin{split}
    \label{eq: PGT1sf}
        a(t) &= \frac{A}{B} \sqrt{\Omega_{r,0}} \tau+
        \frac{\Omega_{m,0}(3B^{2}-1)A^{2}}{16 B r^{2}}\tau^{2}
        \\
        &+\frac{\Omega_{m,0}^{3}(27B-121)A^{4} (B^{2}-1)}{49152 B^{4} \Omega_{r,0}}\tau^{3}
        \\
        &+\frac{\Omega_{m,0}^{3}(27 B^{2}-121)A^{4}(B^{2}-1)}{49152 B^{4}\Omega_{r,0}}\tau^{4}
        \\
        & +\left[ \dfrac{\splitdfrac{\splitdfrac{
       (-441\Omega_{m,0}^{4}+98304 B^{2} \Omega_{\Lambda,0} \Omega_{m,0}^{4}}
       {+ 1421 B^{2}\Omega_{m,0}^{4}+32768\Omega_{\Lambda,0} \Omega_{r,0}^{3}}}
       {-980\Omega_{m,0}^{4})A^{5}}}
       {1310720 B^{5}}\right] \tau^{5}
       \\
       &+O(\tau^{6})
\end{split}
\end{equation}
%
%%%%%%%%%%%%%%%%%%%%%%%%%%%%%%%%%

%%%%%%%%%%%%%%%%%%%%%%%%%%
where $\tau$ is the proper time and
\begin{equation}
A=\frac{-4}{3(3b_{3}-2b_{1}-b_{2})},
\end{equation}
is a combination of coefficients in the Lagrangian term \eqref{eq: Lagt} and
\begin{equation}
B=\frac{6a_{1}+a_{2}+a_{3}+4a_{5}-2a_{6}}{3b_{3}-2b_{1}-b_{2}}\Psi_{r},
\end{equation}
is a combination of the radiation component of the torsion function $\Psi$ in \eqref{eq: Torpg} and coefficients in the Lagrangian term \eqref{eq: Lagr}.
Figures (\ref{fig: PGT1Dc}) and (\ref{fig: PGT1Da}) show the evolution of line-of-sight comoving distance $d_c$ and angular diameter distance $d_A$ for this case. The values of parameter chosen for the plots are $B=0.8$ and $B=1.2$ and $A=\frac{4}{3}$ while the density parameters coincide with flat $\Lambda$CDM data.
Another Class of PGT referred in \cite{Barker} as $^{3}E$, allows for a cyclic universe where the scale factor is given by
\begin{equation}
\label{eq: PGT2sf}
    a(t) = c_1 (\cosh{(c_2 t)}-1),
\end{equation}
where
\begin{equation}
    c_1 = \frac{\Omega_{r,0}}{\Omega_{m,0}},\\ \nonumber
    \end{equation}
\begin{equation}
\begin{split}
    c_2 =
    &
    {(\frac{3}{2}a_1+\frac{1}{2}(a_{2}+a_{3}+a_{6})+a_{5})(3b_{3}-2b_{1}-b_{2})}
    \over\displaystyle{
    (\frac{1}{2}(3a_{1}+a_{2}+a_{3}+a_{6})+a_{5})^2
      \atop\quad
    {}-4(\frac{1}{4}(6a_{1}+a_{2}+a_{3})+a_{5}-\frac{1}{2}a_{6})^2}
\end{split}
\end{equation}

\begin{figure}[h!]
    %\centering
    \includegraphics[scale=0.4]{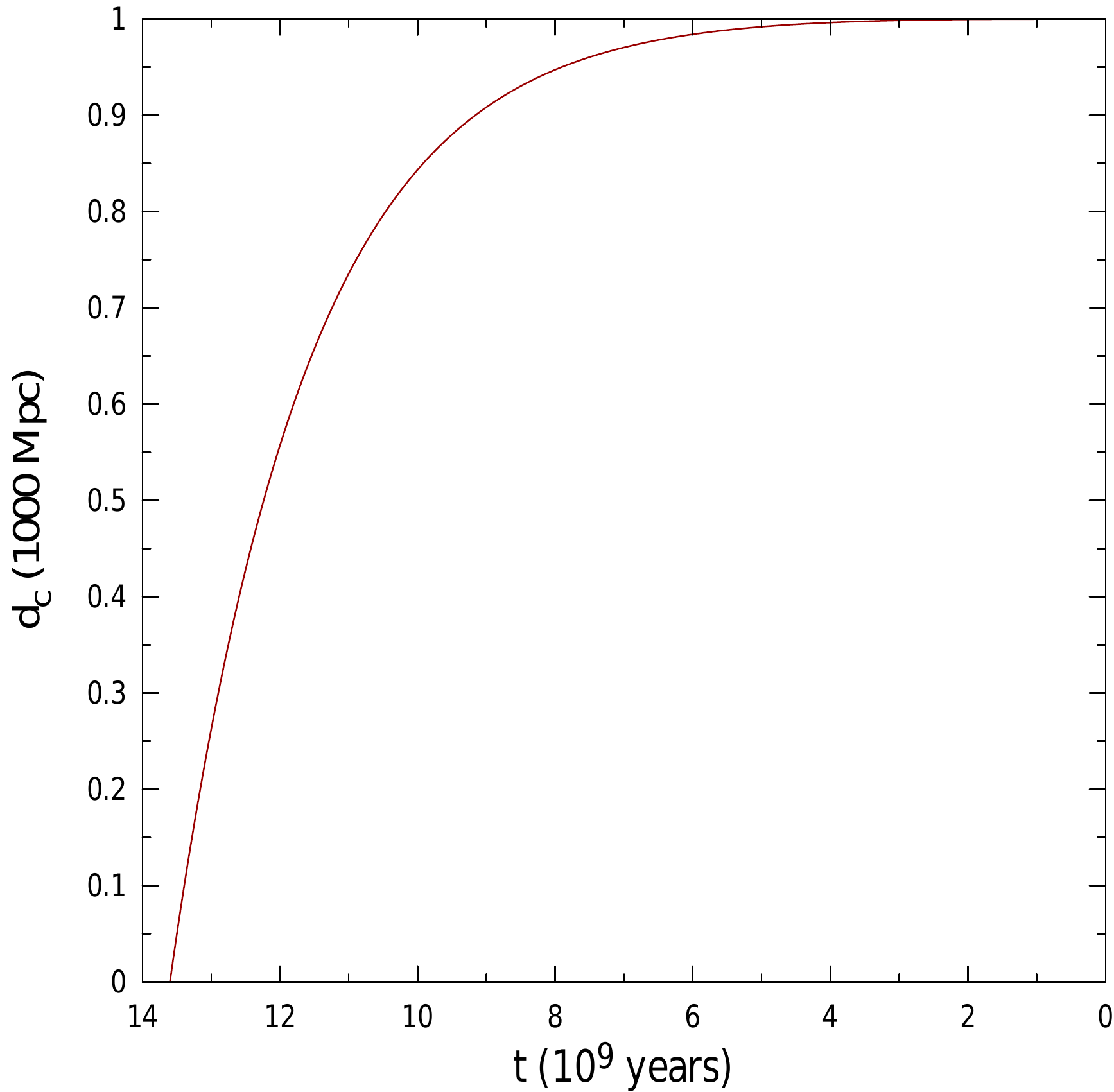}
    \caption{The evolution of line-of-sight comoving distance versus the cosmic time in Poincar{\'e} gauge theory of gravity for the scale factor given by equation \eqref{eq: PGT2sf}. The present day time is set to $t_0=0$ in this plot and the time axis unit is in Giga Years so the horizontal axis show the age of the universe.}
    \label{fig: PGT2Dc}
\end{figure}
\begin{figure}[h!]
    \centering
    \includegraphics[scale=0.4]{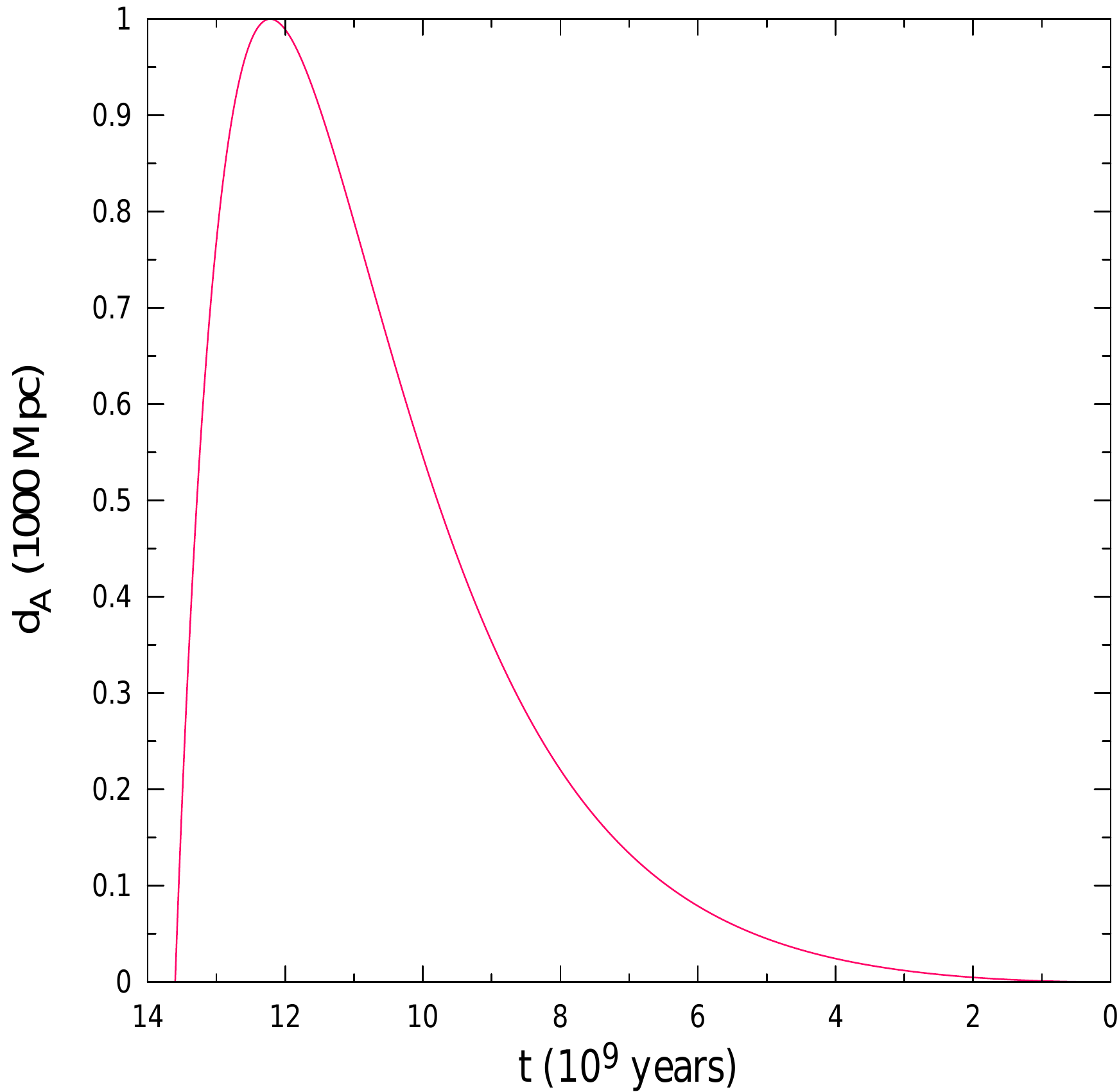}
    \caption{The evolution of angular diameter distance versus the cosmic time in Poincar{\'e} gauge theory of gravity for the scale factor given by equation \eqref{eq: PGT2sf}. The present day time is set to $t_0=0$ in this plot and the time axis unit is in Giga Years so the horizontal axis show the age of the universe.}
    \label{fig: PGT2Da}
\end{figure}

Figures (\ref{fig: PGT2Dc}) and (\ref{fig: PGT2Da}) show the evolution of of line-of-sight comoving distance $d_c$ and angular diameter distance $d_A$ for this case. Note that for both solutions of PGT theories discussed here, a suitable choice of parameters can change the time delay distance in \eqref{eq: tdd} to change in such a way that the measured Hubble parameter be lower than its GR value, similar to the results in table (\ref{table:time_delays}) for Einstein-Cartan theory.

%%%%%%%%%%%%%%%%%%%%%%%%%%%%%%%%%%%%%%%%%
\section{Statements and Declarations}
\subsection{Funding}
No funding was received for conducting this study.
\subsection{Competing interests}
The authors have no relevant financial or non-financial interests to disclose.
\subsection{Data Availability}
Data sharing not applicable to this article as no datasets were generated or analysed during the current study.
\subsection{Author contribution statement}
Both authors contributed to the study conception and design. Both authors contributed equally to the calculations. The final manuscript text was written by Siamak Akhshabi. Saboura Zamani prepared all the figures. Both authors read and approved the final manuscript.
%%%%%%%%%%%%%%%%%%%%%%%%%%%%%%%%%%%%%%%%%

\end{document}